# High sensitive MIS structures with silicon nanocrystals grown via solid state dewetting of silicon-on-insulator for solar cell and photodetector applications

Mansour Aouassa[1*]; Saud A. Algarni[2]; Ibrahim O. Althobaiti[3]; Luc Favre[4] and Isabelle Berbezier[4].

[1]Department of Physics, College of Science and Arts, Jouf University, Al-Qurayat Branch, P.O. Box 756, Saudi Arabia.
[2]Department of Physics, College of Science, Taif University, P.O. Box 11099, Taif 21944, Saudi Arabia
[3]Department of Chemistry, College of Science and Arts, Jouf University, Al-Qurayat Branch, P.O. Box 756, Saudi Arabia.
[4]IM2NP, CNRS, Aix Marseille Université, Université Toulon, 13397 Marseille, France.

[*]Corresponding author: maouassa@ju.edu.sa

**Abstract**

This work reports an original method for the fabrication of Metal-Isulator-Semiconductor (MIS) structures with silicon nanocrystals (Si NCs) based active layers embedded in the insulating $SiO_2$ oxide, for high performance solar cell and photodetector applications. The Si NCs are produced via the in situ solid-state dewetting of ultra-pure amorphous silicon-on-insulator (a-SOI) grown by solid source molecular beam epitaxy (SSMBE). The size and density of Si NCs are precisely tuned by varying the deposited thickness of silicon. The morphological characterization carried out by using atomic force microscopy (AFM) and scanning electron microscopy (SEM) shows that the Si NCs have homogeneous size with well-defined spherical shape and densities up to $\sim 10^{12}$ /$cm^2$ (inversely proportional to the square of nominal a-Si thickness). The structural investigations by high resolution transmission electron microscopy (HR-TEM) show that the ultra-small Si NCs (with mean diameter ~7 nm) are monocrystalline and free of structural defects. The electrical measurements performed by current versus voltage (I-V) and photocurrent spectroscopies on the Si-NCs based MIS structures prove the efficiency of Si NCs to enhance the electrical conduction in MIS structures and to increase (x10 times) the photocurrent (i.e. at bias voltage V = -1 V) via the photo-generation of additional electron-hole pairs in the MIS structures. These results evidence that the Si NCs obtained by the combination of MBE growth and solid-state dewetting are perfectly suitable for the development of novel high performance optoelectronic devices compatible with the CMOS technology.

**Introduction**

The development of new processes to produce semiconductor nanostructures and their integration in many advanced fields such as nanocrystals in solar cells, quantum wells in lasers and nano-wires in FinFet transistors, has generated a huge expansion of this field of research which has find an increasing number of potential applications [1-6]. Among the most mature NCs studied at the nano-scale, the best controlled are undoubtedly the silicon (Si) ones (with 60 years of permanent research devoted to this semiconductor material since the beginning of microelectronics [7, 8]). In this context, Si NCs have been the subject of intense research since their discovery by Wagner and Ellis in 1964 and their integration into functional devices is increasingly frequent [9].

Their use is generally subject to the control of their electronic and structural properties but also to the production strategy to ensure the high quality and reliability of these nanostructures, for example the Si NCs must have a high density and good size homogeneity to minimize the dispersion of their physical properties [10]. Indeed, if the intrinsic chemical purity is essential because it guarantees the performance of the devices, a high density of silicon NCs without structural defects with a controllable size is also crucial in order to ensure good sensitivity of the optoelectronic devices. In addition, the use of substrates compatible with CMOS technology (Silicon, $SiO_2$, Germanium ...) is another major issue for the diversification of fields of application and the reduction of production costs.

Many works have thus been devoted to the study and integration of Si NCs in Metal-Insulator-Semiconductor (MIS) structures for the development of photodetectors, non-volatile memories and solar cells combining their high yields and the low cost of Si [11-14]. Many techniques have been used for the fabrication of Si NCs, the chemical vapor deposition (CVD), the annealing of Si rich oxide (SRO) layers or the implantation of Si in $SiO_2$ layers. However, these techniques commonly produce NCs with crystalline defects and/or impurities (specially at the NCs/$SiO_2$ interface), inhomogeneous size, low densities and imprecise positioning [15-20] which make them unsuitable for optoelectronic applications.

In this paper we propose a technique based on the coupling between MBE growth and in-situ solid state dewetting for the fabrication of high-quality Si NCs on an ultrathin thermal $SiO_2$ layer itself on Si (001) substrate (SOI). Dewetting is a phenomenon where a thin film ruptures on a substrate, which leads to droplets formation. We show that the solid state dewetting technique developed [6; 11; 16; 21] allows the fabrication of ultra-pure, homogeneous, highly dense Si NCs with tunable size at nanometer scale which gives an effective control of their band gap energy (via the energy quantification of carriers) and of their optical and electronic properties [11]. These advantages highlight that the dewetting process is very mature for the growth of Si NCs compatible with CMOS technology.

**Experimental**

The fabrication of the MIS solar cells studied in this work is based on two steps, first the growth of amorphous SOI via deposition of amorphous Si (a-Si) thin films in an ultra-high vacuum (UHV) MBE chamber on a silicon di-oxide ($SiO_2$) film and second the solid state dewetting / crystallization of a-Si. The fabrication process includes the following steps: 1) growth of a thin $SiO_2$ film with a thickness of 5 nm by rapid thermal oxidation (RTO) of Si (001) substrate (JETFIRST 8inch Jipelec); 2) Deposition of 1 nm and 2 nm a-Si at room temperature (RT) on the $SiO_2$/Si (001) substrates. A RIBER 32 UHV-MBE chamber with residual pressure of $10^{-10}$ torr equipped with an electron gun was used for the evaporation of ultra pure silicon; 3) Transformation of the thin a-Si film, into 3D Si NCs by annealing at 730°C for 30 min; 4) Deposition of 1 nm a-Si at RT to encapsulate the NCs and avoid contamination during the air exposure ; 5) Ex situ capping of the Si NCs with 45 nm SiO$_2$ deposited by plasma enhanced chemical vapor deposition (PECVD) followed by annealing at 800°C for 5 min, to improve the structural quality of the Si NCs/SiO$_2$ interface which plays an important role in the electrical conduction mechanisms. 6) Deposition of transparent 15 nm AuPd film by compact plasma sputter coater (Oxford) to create the metallic contacts and obtain MIS solar cells (photodetectors). The fabrication steps of these MIS structures are summarized in the figure (1).

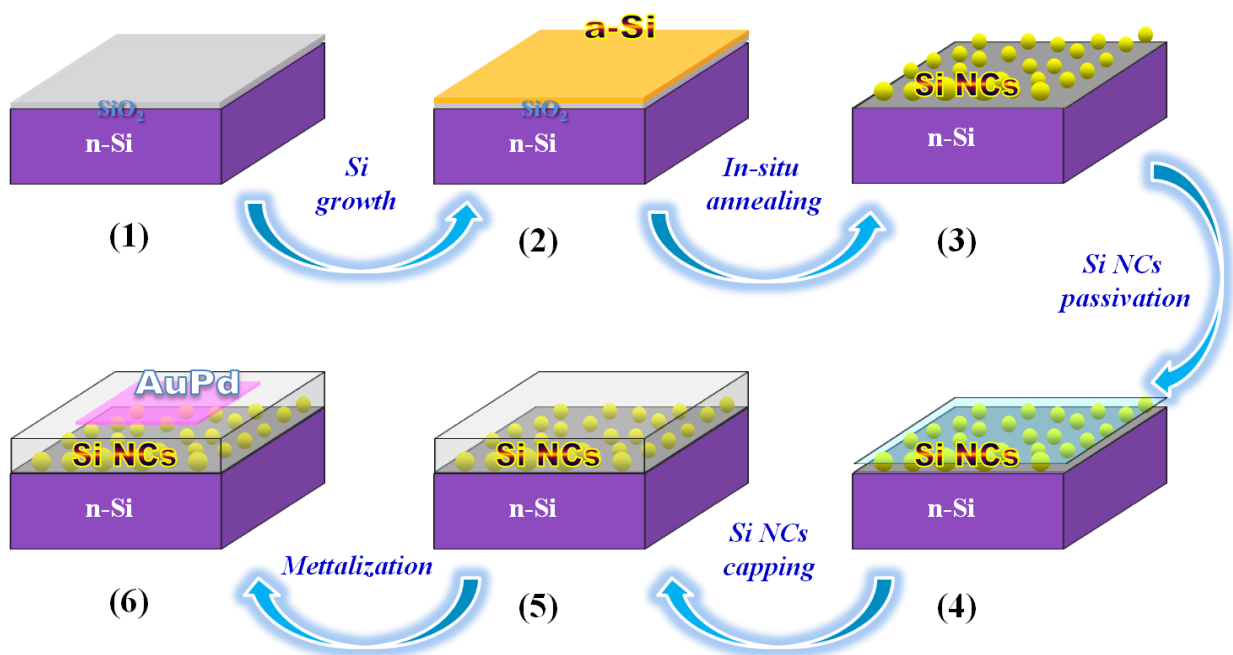

*Fig. 1: The fabrication steps of these Si NCs-MIS structures*

The morphological and structural properties of the fabricated NCs are studied by atomic force microscopy (AFM) in air (type XE-100 from Park Systems Company) operated in tapping mode, by scanning electron microscopy (SEM) Lyra-Tescan and by high resolution transmission electron microscopy (HR-TEM) type JEOL 2010F operated at 200 KV.

The electrical transport in MIS devices was investigated using HP 4140B Pico-Ammeter DC Voltage Source (controlled by Labview program). The maximum and minimum values of the applied voltage, the number of acquisition points for each voltage and the incremental step are properly chosen. The current-voltage characteristics (I-V) are recorded in the dark and under polychromatic illumination using a tungsten halogen lamp.

The photocurrent (PC) equipment includes a light source (Tungsten Halogen lamp) and a monochromator (from Spectra Physics Company). A semi reflective mirror separates the beam from the monochromator into two parts. One part illuminates the reference photodiode and the other part is modulated at 60 Hz to illuminate the sample on the side of the transparent AuPd electrode. The photodiode measures the light energy received by the sample. The photocurrent measurements are carried out by a Lock-In Amplifier (EG&G Princeton Applied Research 5208) to overcome the noise due to stray light.

**Results**

The Si-NCs obtained by deposition/dewetting have been observed by AFM. The images show randomly distributed NCs with homogeneous size and densities $D(\tau) \approx 10^{12}/cm^2$ and $D(\tau) \approx 10^{11}/cm^2$ for deposited thickness of 1 nm and 2 nm respectively. Such densities are higher than those reported when using other deposition methods such as CVD, or PECVD. In addition, these NCs are characterized by well-defined hemispherical shape as shown in Fig.2-a.

Another advantage of the fabrication method is the easy and high precision control / tuning of the NCs lateral size (Φ) by varying the thickness (τ) of the deposited a-Si layer. The empirical evolution of Φ and $D(\tau)$ with $\tau$ deduced from the experimental results is given by (Fig. 2b) respectively:

$$\Phi = \alpha\, \tau$$

With α is a constant = 7.

$$D(\tau) = \frac{5.6}{\tau^2} \times 10^{11}\, /cm^2, \quad (\tau \text{ in nanometer})$$

It gives a density of $5.6 \times 10^{11}/cm^2$ and $1.4 \times 10^{11}/cm^2$ for 1 nm and 2 nm a-Si deposited thickness respectively, in good agreement with the densities measured experimentally.

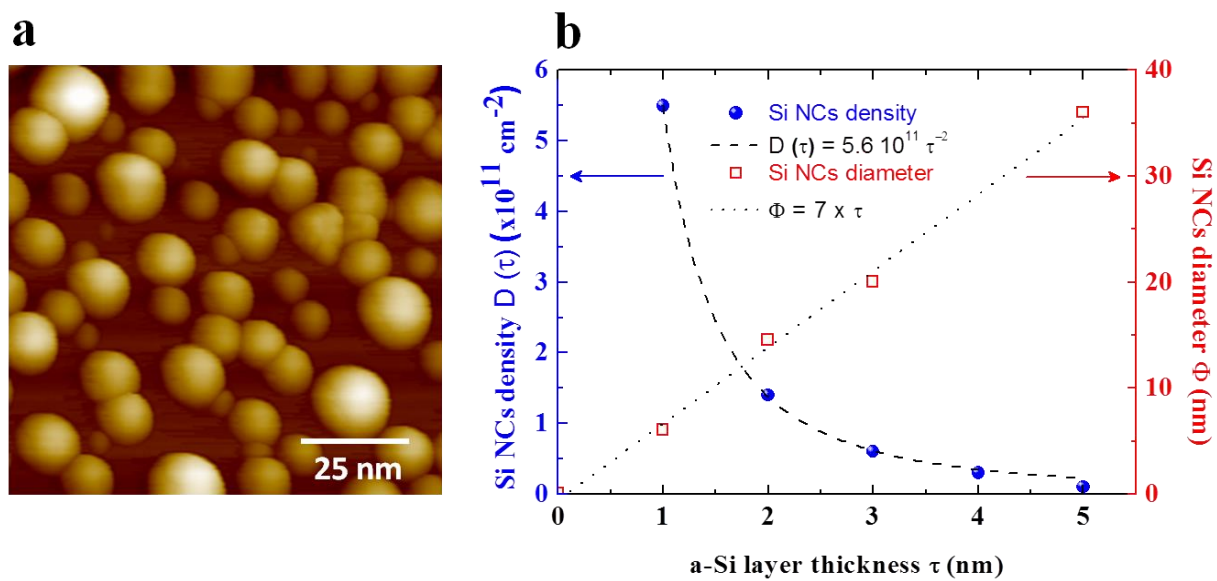

*Fig.2: (a) Typical AFM image of Si NCs (τ =2 nm) obtained by in-situ solid state dewetting of a-Si; (b): Evolution of Φ and D with τ.*

The study of the morphological and structural properties carried out by SEM confirms the AFM results (Fig. 3a-b). The very good size homogeneity and large density of NCs can be well appreciated for 1 nm a-Si (Fig. 3a) while size inhomogeneities start to appear for 2 nm a-Si (Fig. 3b). In addition, TEM cross-section observations show that the ultra-small Si NCs are monocrystalline and free from crystalline defects (Fig. 3c).

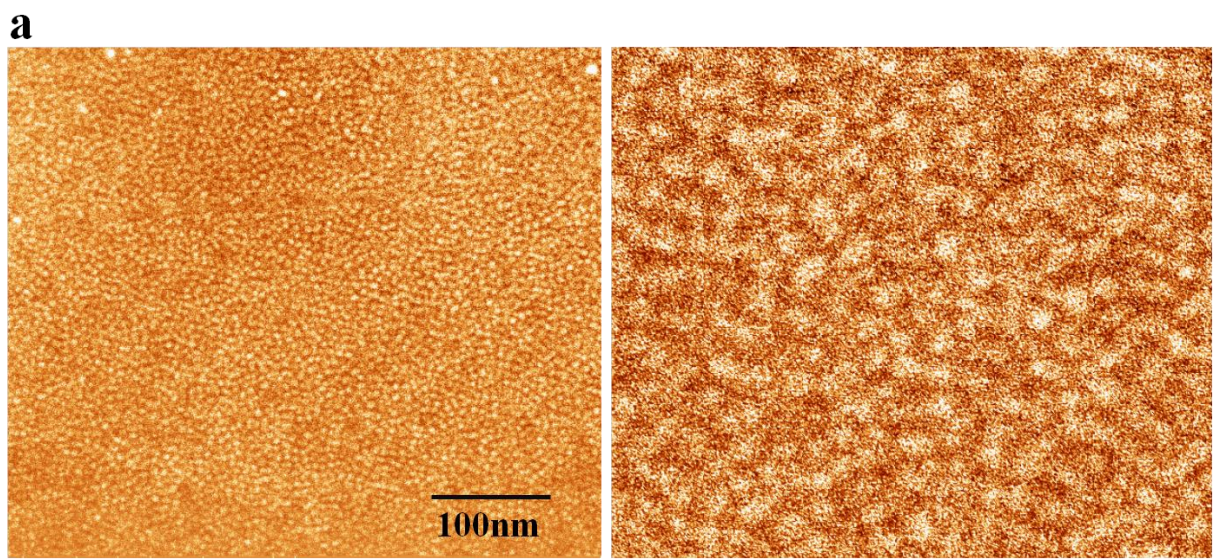

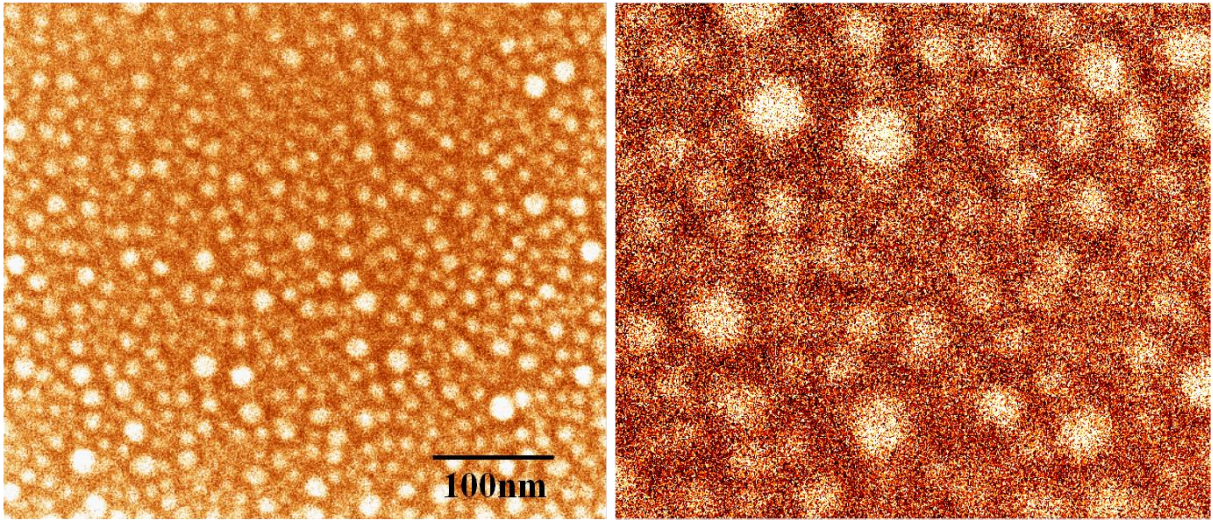

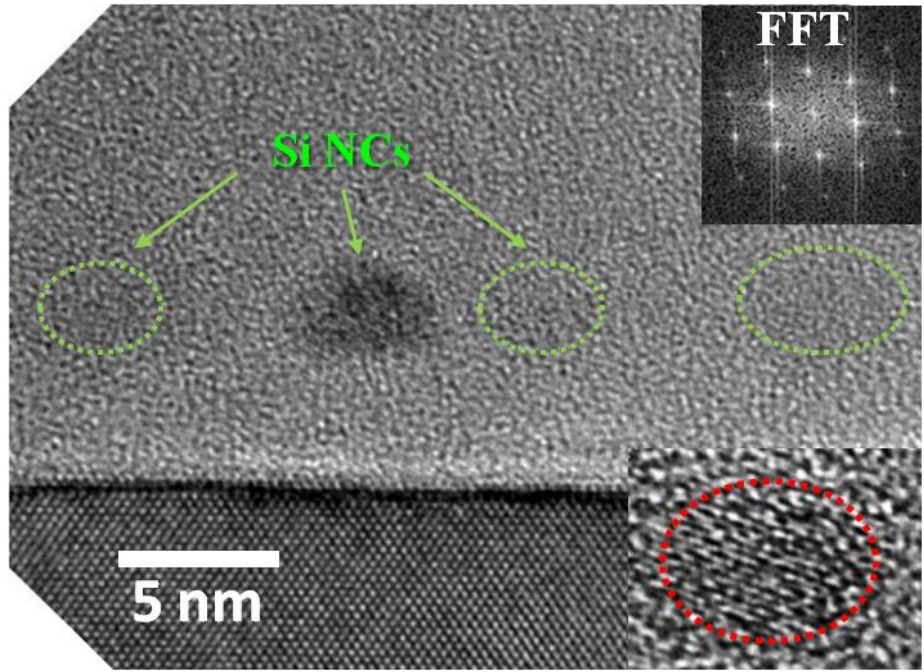

*Fig.3: (a) and (b) SEM images of Si NCs obtained by dewetting of 1 nm and 2 nm thick a-Si showing the homogenous size of the NCs; (c) TEM cross-section image of a single NC obtained from 1 nm thick a-Si. In the insets are given the FFT of the image and a zoom of the NC.*

All these results prove that the fabrication method is able to produce Si NCs fully embedded in a $SiO_2$ insulator layer and that meet the physical criteria required by the optoelectronic industry.

In order to evaluate the influence of the Si NCs inserted in the MIS structure, on the electrical transport properties, three MIS solar cells devices were fabricated using first only the oxide without Si NCs as reference (MIS_Ref.) and second with Si NCs of two different sizes embedded in the oxide (Fig. 4): 1) the ultra-small NCs with $\Phi$ = 7 nm (MIS_1) and 2) the larger NCs with $\Phi$ = 14 nm (MIS_2).

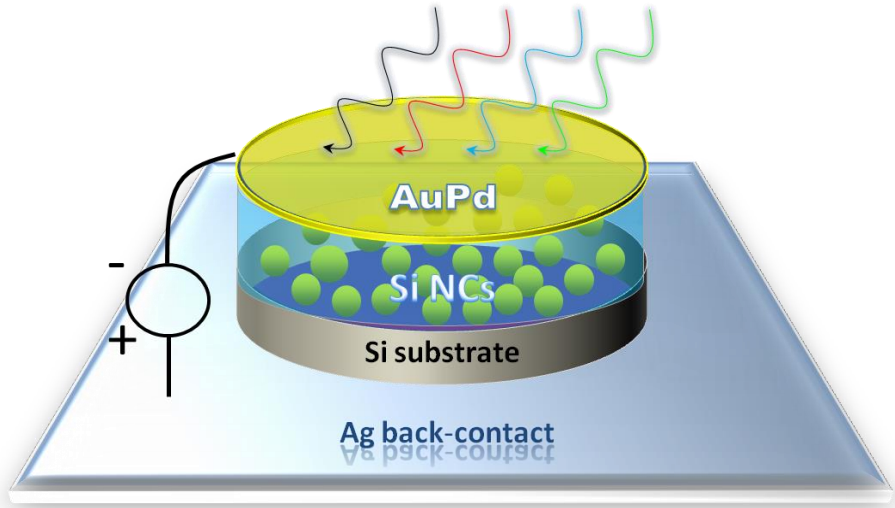

*Figure 4: Schematic structure of MIS solar cell with Si NCs*

We performed I-V measurements on the three MIS structures in the dark at room temperature. Fig. 5 shows the I-V characteristic of MIS_2 structure; this characteristic is typical of an asymmetric N-type MIS structure with a very low inversion current and a very large accumulation current. This shows that the transport is governed by the oxide despite the presence of NCs. To better understand the conductance modes involved in the electrical transport, we can correlate the I-V behavior with the static current transport model of De Salvo schematized in Fig. 5b [12]. For a very low positive voltage, the electrons can travel from the substrate to the NCs via the tunnel conduction mode (mode I). By increasing the voltage beyond $V_g \geq 2$ V, the potential barrier created by the control oxide (45 nm) becomes triangular (mode II), in this case the probability of passage of electrons from the Si NCs to the metal grid via the Fowler-Nordheim conduction mode increases, this increase leads to a significant increase in electrical current. For a very high voltage, the potential barrier height dramatically decreases and the electrons can overcome the potential barrier and the electrical current increases further (mode III and IV). For a negative tension, we observe a very low inversion electrical current compared to the accumulation current, this phenomenon is mainly related to the doping of the substrate (type N in this study) and to the asymmetry of the MIS structure, i.e. there is no conduction by tunnel mode from the gate to the Si NCs since the control oxide is too thick (45 nm).

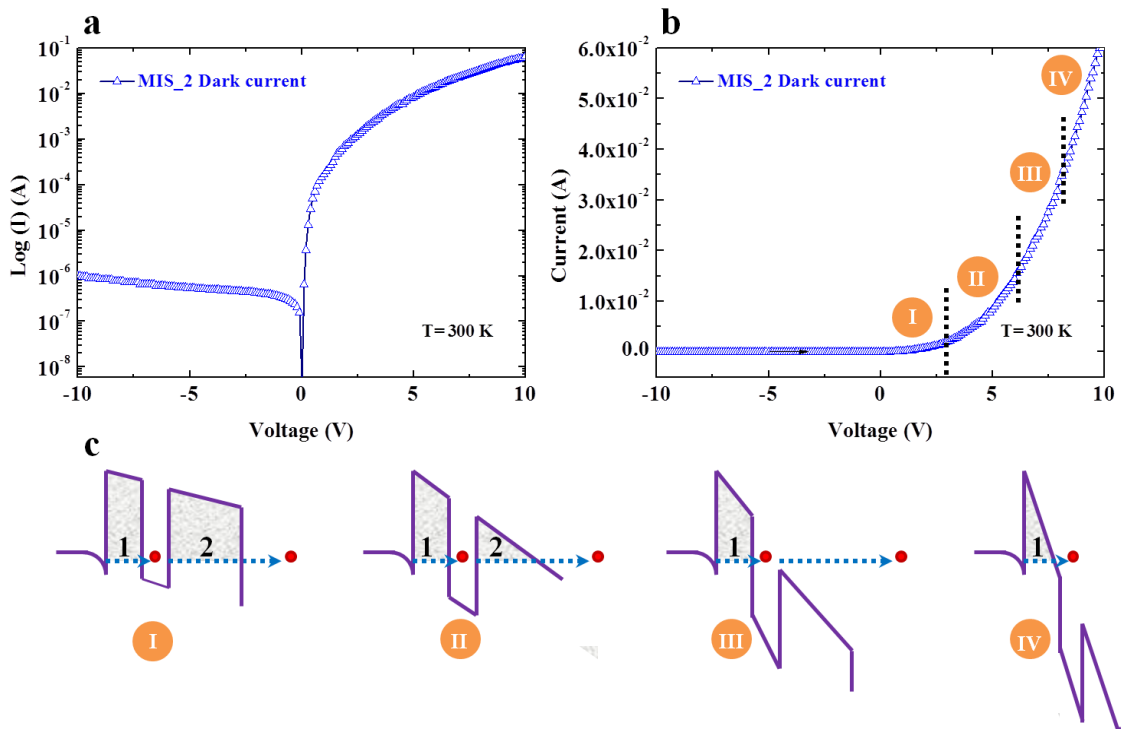

*Figure 5: The variation of current as a function of gate voltage (I-V) for the structures Si NCs MIS_2 plotted in logarithmic (a) and linear (b) scales. Energy band diagrams for different grid voltages applied to the MIS structures with Si NCs (c).*

Fig.6a compares the I-V characteristics of the different MIS structures measured in the dark at room temperature without Si NCs (MIS_REF) and with the two sizes of Si NCs. It is clear that the current increases in the presence of Si NCs inserted in the oxide layer of MIS structures for forward or reverse bias. This increase by two-three orders of magnitude is attributed to the electrical relay created by the NCs in the oxide as explained in Fig.6b [12]. It is also observed that the current increases with the Si NCs diameter because the distance traversed by the carriers in the silicon increases compared to the distance traversed in the oxide during the passage of electrons from the silicon substrate to the grid.

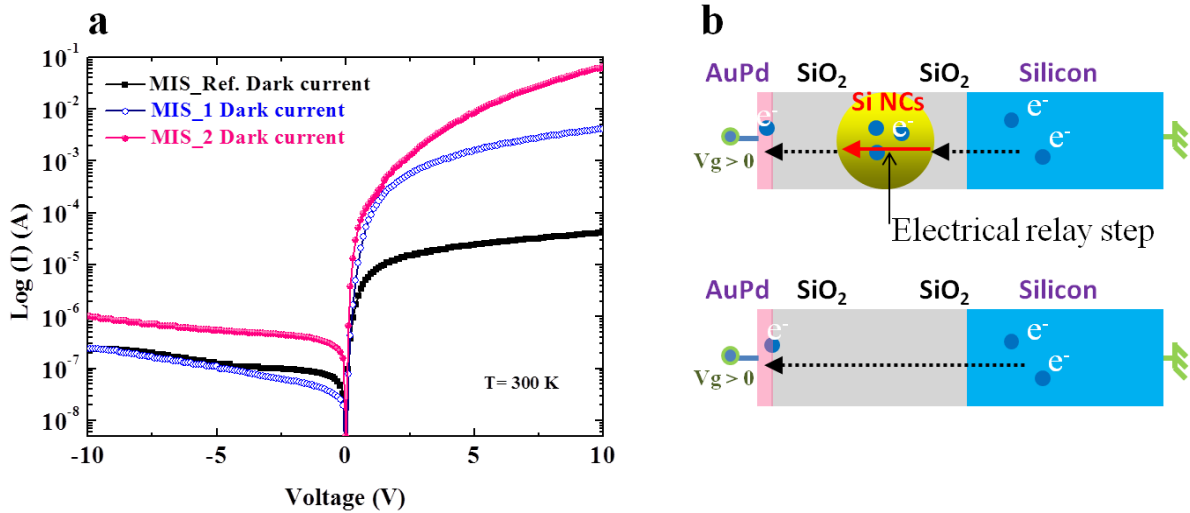

*Fig. 6: Variation of the Log of the current as a function of the gate voltage for the reference structure and the structures with Si NCs highlights the influence of the NCs on the I-V characteristic (a). Diagram explains the effect of NCs on the phenomenon of conduction (b).*

Fig. 7 shows the photocurrent versus voltage (I-V) characteristic of the three MIS structures excited by white light at room temperature. A significant increase of the photocurrent is observed when the MIS structures contain Si NCs. For forward bias Vg= 1 V, the photocurrent of the Si MIS structures increases ~60times and more than 300 times for MIS_1 and MIS_2 as compared to MIS_Ref respectively.

For a reverse bias Vg= -1 V, a photocurrent increase of 5 times and 10 times is observed for MIS_1 and MIS_2 structures respectively as compared to the reference sample (MIS_Ref) (increases from 200 µA to 120 µA) and increases 10 times for the MIS _2 structure (increases from to 200 µA to 120 µA).

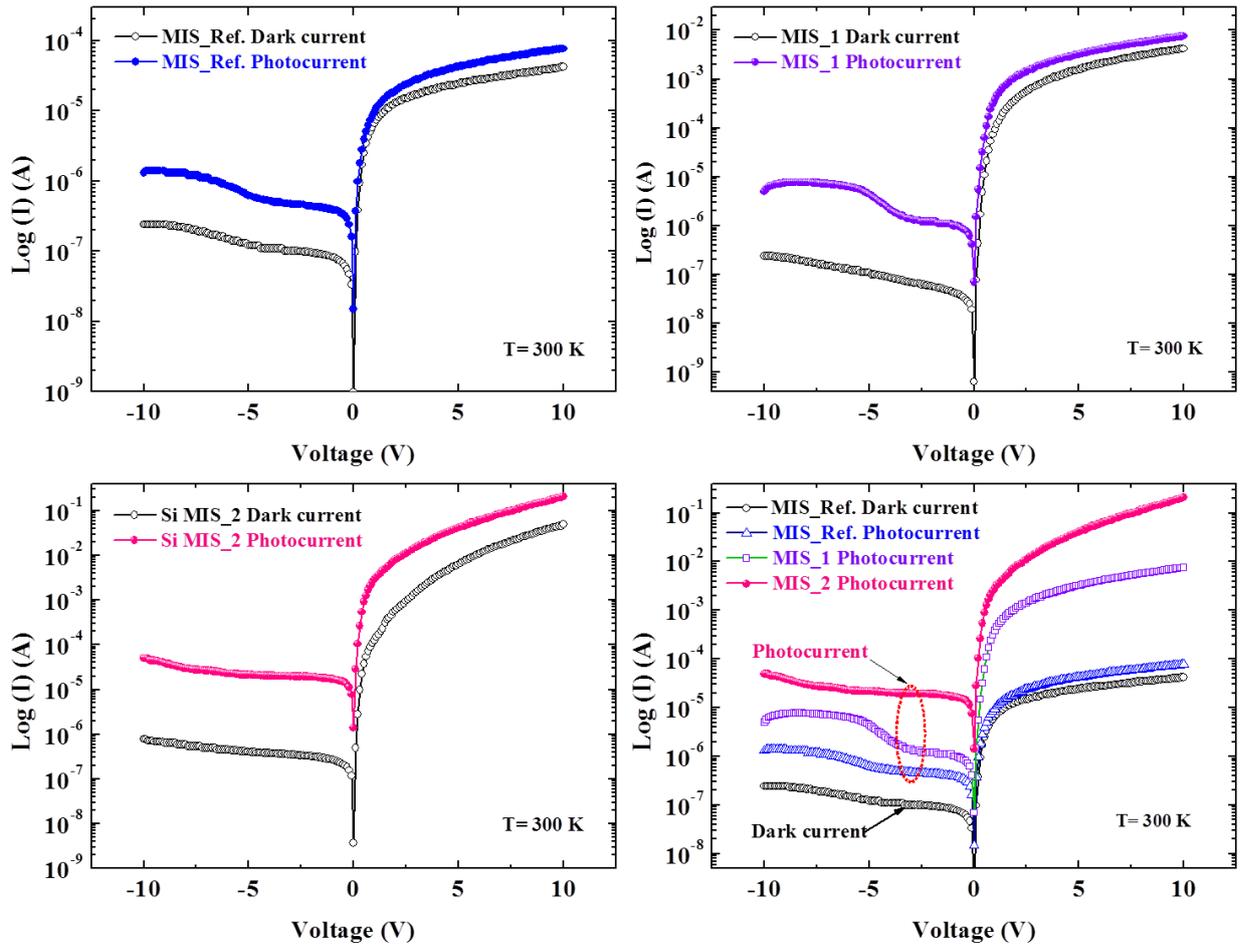

*Fig. 7: Comparison between the I-V characteristics measured in the dark (black curve, dot symbols) and under white light illumination for the three MIS structures: without (Ref) and with Si NCs 7 nm (violet curve) and 14 nm (pink curve).*

A such large photocurrent improvement is attributed to the efficient contribution of Si NCs to the photo-generation of electron-hole pairs in MIS structures as explained in the Fig.8 which shows the energy band diagram of the MIS structure with nanocrystals including the interband optical transitions responsible for the photoresponse of the nanocrystals and the substrate, the violet arrows represent the electrons transport through the Si NCs. This large current amplification is higher than that ever observed when the Si NCs are fabricated either by Si implantation, or by deposition processes (plasma or CVD) and by decomposition of $SiO_x$ films during annealing [21-23]. This higher photocurrent is attributed to the higher structural properties and surface passivation of Si NCs. In particular, we suggest that the in situ deposition of an a-Si encapsulation layer reduces the density of states at the Si NCs-SiO$_2$ interface and consequently improves the electrical transport in the MIS structure.

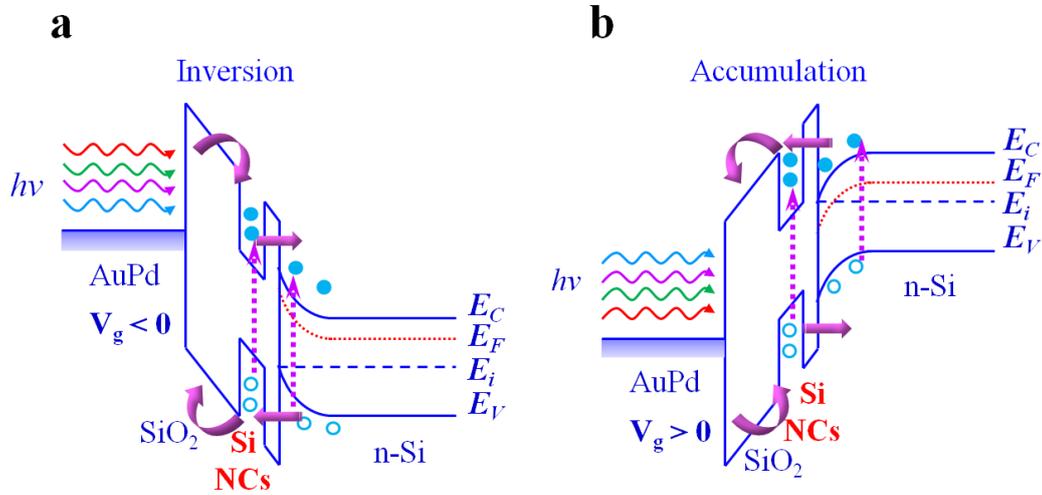

*Fig. 8: Schematic representation of the energy band diagram of a Si NCs based MIS structure in (a) reverse and (b) direct polarizations.*

To better understand the photocurrent efficiency, we have selected the structure containing the smallest NCs (MIS_1) in which a strong quantum confinement can be expected and reference structure without nanocrystals (MIS_Ref.). We have plotted the spectral responses of the photocurrent emission in the visible range at room temperature of MIS_Ref and MIS_1 (Figure 9). The photocurrent spectrum of MIS_1 exhibits high photocurrent intensity on a wide wavelength range (between 500 nm and 960 nm) compared to MIS_Ref. with a record high responsivity of 3.2 A/W at 698 nm. The additional photocurrent observed for the MIS_1 comes from the absorption of photons in the Si nanocrystals. The peak observed at 1024 nm (1.2 eV) for the two samples (with / without Si nanocrystals) is attributed to an absorption linked to the gap of the silicon substrate at the interface Si/SiO$_2$.

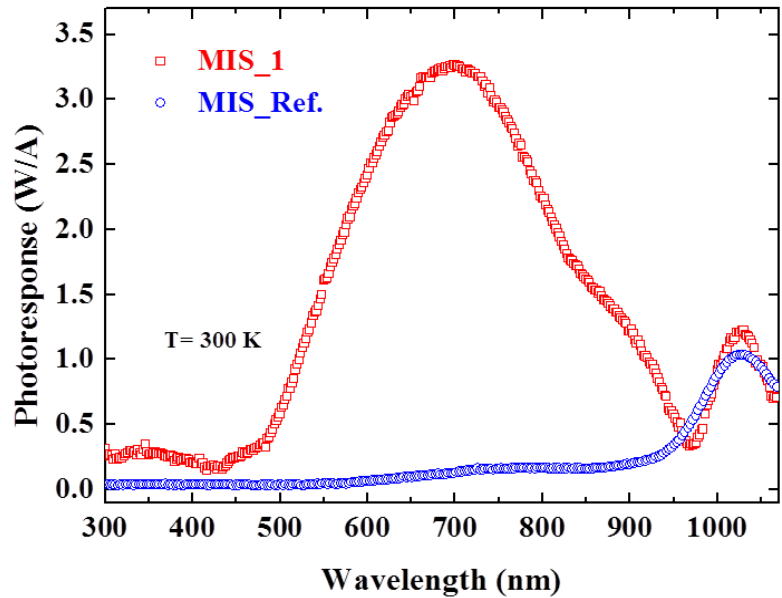

*Fig.9: spectral response of the photocurrent emission of Si MIS_1*

**Conclusion**

In this work we have reported an efficient process for the fabrication of high sensitivity MIS photodetector. It is based on the monolithic integration of ultra-pure, defect-free and very dense (up to ~ $10^{12}/cm^2$) Si NCs grown via solid state dewetting of silicon-on-insulator into the MIS structures. The morphological and structural results are correlated to the current-voltage behavior and photocurrent measurements. The results evidence a strong increase of the photocurrent about three orders of magnitude under inverse polarization. The Si NCs are also used for the fabrication of a highly sensitive photodetector in a wide wave-length range, up to 3.2 A/W in the near infra-red region (at ~700 nm). These results show that the simple large scale two steps process developed produces ultra-small Si NCs that are perfectly compatible with CMOS technology and very promising for the development of high performance opto-electric devices.

**Research Data Policy:** Data sharing and data citation is encouraged.
**Data Availability Statements:** The datasets generated during and/or analysed during the current study are available from the corresponding author on reasonable request.
**Acknowledgements:** This work was funded by the Deanship of Scientific Research at Jouf University under grant No (DSR-2021-03-03159).